\documentclass[11pt,english]{article}
\usepackage[T1]{fontenc}
\usepackage[latin9]{inputenc}
\usepackage{geometry}
\usepackage{fancyhdr}
\usepackage{graphicx}

\makeatletter
\usepackage{lastpage} % to be able to display page n of N
\makeatletter
\let\ps@plain\ps@fancy % Plain page style = fancy page style
\makeatother
\usepackage{hyperref}
\makeatother

\usepackage{babel}
\usepackage{listings}
\begin{document}

\title{\textbf{Dafny: Statically Verifying Functional Correctness}}

\author{Rachel Gauci\\
{\small{}The University of Edinburgh}\\
{\small{}\href{mailto:s1402642@sms.ed.ac.uk}{s1402642@sms.ed.ac.uk}}}
\maketitle
\begin{abstract}
This report presents the Dafny language and verifier, with a focus
on describing the main features of the language, including pre- and
postconditions, assertions, loop invariants, termination metrics,
quantifiers, predicates and frames. Examples of Dafny code are provided
to illustrate the use of each feature, and an overview of how Dafny
translates programming code into a mathematical proof of functional
verification is presented. The report also includes references to
useful resources on Dafny, with mentions of related works in the domain
of specification languages.

\textit{The research work disclosed in this publication is funded
by the MASTER it! Scholarship Scheme (Malta). The scholarship is part-financed
by the European Union - European Social Fund (ESF) under Operational
Programme II - Cohesion Policy 2007-2013, ``Empowering People for
More Jobs and a Better Quality of Life''.}
\end{abstract}

\section{Introduction}

Dafny is both a programming language and a verifier, capable of performing
full verification of the functional correctness of a program \cite{leino2010autoprogver}.
Verification is possible due to language-specific features such as
pre- and post-conditions, loop invariants, assertions and framing.
The Dafny verifier takes care of proving that the code does indeed
match its annotations, and thus the burden of writing bug-free code
is lifted into that of writing bug-free annotations, with the underlying
assumption that annotations are less prone to errors since they are
shorter and more direct \cite{koenig2012gettingstarted}.

The content of this report is structured as follows. Section \ref{sec:Context}
highlights the motivation and necessity for bug-free code and functional
verification, while Section \ref{sec:Dafny} is fully dedicated to
the Dafny language and verifier. Section \ref{sub:Language-Features}
describes the main language features which enable the functional verification
to occur. Each feature is illustrated with an example written in the
Dafny language, where all examples have been created using Dafny version
1.9.1%
\footnote{This is the latest version of Dafny, which was released only recently
on 22 October 2014, and can be downloaded from CodePlex at \url{http://dafny.codeplex.com/}
\cite{codeplex}.%
} in the Microsoft Visual Studio Professional 2013 integrated development
environment. In Section \ref{sub:From-Code-to}, the report touches
up on how Dafny goes about generating a mathematical proof of functional
correctness from the actual Dafny code. Section \ref{sec:Resources}
is more literature-oriented, with references to good resources for
getting started with Dafny and and a mention of similar specification
languages. Section \ref{sec:Conclusion} then concludes the report
by reiterating over the main presented points and summarizing Dafny's
advantages and limitations.

\section{Context\label{sec:Context}}

Different software systems perform different tasks and have different
requirements, but the one underlying, common requirement is that of
having bug-free and functionally-correct code. However, history has
shown us that this is not an easy requirement to meet. As can be seen
from some of the software failures listed at \cite{softFailures},
even systems which have been developed by highly-skilled developers
and undergone rigorous testing, such as those at NASA and Microsoft,
are in no way immune to bugs.

Functional correctness of a software system is not an easy thing to
prove. With testing methodologies such as unit testing, for example,
we would need to make sure that the test cases cover all possible
paths of execution. Not only is this a huge, and sometimes impossible,
task to undertake, but the testing code itself is also prone to errors.
While this kind of testing does indeed help in the identification
of bugs, it cannot prove that a program is fully functionally correct,
as a single, overlooked input could possibly cause a run-time error
and bring the entire system down.

Traditionally, full verification has been carried out through the
construction of manual proofs or proof assistants which require user
interaction \cite{leino2010autoprogver}, but we would like this process
to be as automated as possible, to reduce both the effort required
and the possibility of errors in the verification stage. This kind
of automatic verification can be done with Dafny, which is both a
language and verifier, and is able to prove full functional correctness
in an automated fashion, while always taking all possible paths of
execution into consideration.

\section{Dafny\label{sec:Dafny}}

The Dafny language relies on high-level code annotations, which must
be explicitly written up by the developer, and the verifier then takes
care of checking that the code does indeed match the annotations.
Not only this, but Dafny also proves code termination in the case
of loops and recursion, and the absence of run-time errors in general,
``such as index out of bounds, null dereferences, division by zero,
etc.'' \cite{koenig2012gettingstarted}. Dafny is therefore capable
of giving a very strong guarantee regarding the functional correctness
of a piece of annotated code.

The Dafny verifier performs static verification of the code, and code
which is not verifiable will not compile. Conversely, compiled code
is always guaranteed to have passed the verification stage. Perhaps
one of the best features of Dafny is that the developer is only required
to produce the correct annotations. Once the annotations are in place,
no user interaction is required to generate the proof of functional
correctness.

\subsection{Language Features\label{sub:Language-Features}}

The Dafny language itself is an imperative and sequential, class-based
language, with support for variables, generic types, loops and conditional
statements.

\sloppypar{}The main types are: \texttt{int} for integers,\texttt{
nat} for natural numbers, \texttt{bool} for booleans,\texttt{ set<T>}
and \texttt{seq<T>} for immutable sets and sequences of values of
the generic type \texttt{T}, respectively, \texttt{array<T}>\texttt{,array2<T>,...,array}\texttt{\textit{n}}\texttt{<T>}
for \textit{n}-dimensional arrays, and user-defined classes and inductive
datatypes \cite{koenig2012gettingstarted}. The built-in \texttt{object}
type is a supertype of all class types, and, one of the major features
of the latest release is the added support for \texttt{char} and \texttt{string}
types \cite{releasenotes}.

\subsubsection{Methods, Functions and Function Methods\label{sub:Methods,-Functions-and}}

\textbf{Methods }are one of the basic units of a Dafny program, and
the skeleton for a Dafny method is presented in Listing \ref{lis:Skeleton-for-a}.
The first thing that stands out is the support for annotations such
as \texttt{requires} and \texttt{ensures}. However, I would like to
point out two other, perhaps more subtle, interesting specification
details. Firstly, note that the method allows for multiple return
values (something which is very much to my personal liking as I believe
it can be handy in programming scenarios were one has to resort to
the use of tuples or objects). In Dafny, returning values from a method
involves assigning a value to these output variables before the method
returns. Secondly, the return values are not only typed but also named,
and this allows us to easily refer to them in our annotations.

\begin{lstlisting}[caption={Skeleton for a Dafny method.},label={lis:Skeleton-for-a},float,basicstyle={\small\ttfamily},tabsize=4]
method MethodSkeleton(inParamName: <inParamType>, ...)
		returns (outParamName: <outParamType>,...)
	requires <Precondition>;
	modifies <Frame>;
	ensures <Postcondition>;
	decreases <Rank>;
{
	<MethodBody>
}
\end{lstlisting}
One other important specification detail, which cannot be inferred
from Listing \ref{lis:Skeleton-for-a}, is that the input parameters
are read-only. This is important because Dafny will use preconditions
on these parameters to verify the code for all possible executions
of the program, and therefore all possible parameter values, and this
kind verification would not be possible if the method were free to
change the input parameters at its own will.

The \texttt{<MethodBody>} can be arbitrarily long and it is verified
to make sure that it satisfies all of the method's annotations. However,
when the method is being used from other methods, Dafny ``forgets''
\cite{koenig2012gettingstarted} about its body and considers the
postconditions (denoted by \texttt{ensures}) to be the only thing
it knows about this method. Thus we can see the method annotations
as ``fixing the behavior of the method'' \cite{koenig2012gettingstarted}.
This significantly simplifies the verification process and allows
Dafny ``to operate at reasonable speeds'' \cite{koenig2012gettingstarted}.

\begin{lstlisting}[caption={Skeleton for a Dafny function.},label={lis:functionSkeleton},float,basicstyle={\small\ttfamily},tabsize=4]
function FunctionSkeleton(inParamName: <inParamType>, ...)
		: <returnType>
	requires <Precondition>;
	reads <Frame>;
	ensures <Postcondition>;
	decreases <Rank>;
{
	<UnaryExpression>
}
\end{lstlisting}
As seen in Listing \ref{lis:functionSkeleton}, \textbf{functions}
in Dafny are somewhat similar to methods in that they also can make
use of input parameters and annotations. However, this is where the
similarity ends, as functions in Dafny follow more closely the notion
of a mathematical function: they cannot write to memory and have a
single, unnamed return value. 

Unlike method bodies, the body of a function is \textit{not} forgotten
during verification, and this allows functions to be used directly
in annotations. However, a function body cannot contain more than
one expression (note \texttt{<UnaryExpression>} in Listing \ref{lis:functionSkeleton}
versus \texttt{<MethodBody}> in Listing \ref{lis:Skeleton-for-a}),
and we will come back to this distinction in Section \ref{sub:Assertions},
with code samples of verifiable and non-verifiable annotations.

Functions also introduce the concept of \textbf{ghosting}. Both variables
and functions can be ghosted, and while the Dafny verifier makes no
distinction between regular and ghost fields, ghosts are completely
ignored by the compiler. Functions and other ghosted variables are
therefore only used during the verification stage, and have no affect
on the execution speed or memory space of the final, executable code.
If the code of a function does need to be used by runnable code, we
can easily change it to a \textbf{function method} and it will be
included in the compiled program. In my opinion, ghosting is one of
the nicest features of Dafny, as it allows for static verification
to occur without affecting the performance of the code at run-time.

\subsubsection{Pre- and Postconditions}

Preconditions and postconditions are boolean expressions which are
annotated with \texttt{requires} and \texttt{ensures}, respectively,
and are included as part of a method or function's declaration, as
seen in Listings \ref{lis:Skeleton-for-a} and \ref{lis:functionSkeleton}.
Preconditions must hold \textit{before} method execution, and postconditions
must hold \textit{after} the method returns. As clearly explained
in \cite{leino2010autoprogver}, ``It is the caller's responsibility
to establish the precondition{[}s{]} and the implementation's responsibility
to establish the postcondition{[}s{]}.'' Dafny will make sure the
postconditions hold for all possible invocations of the program, assuming
the preconditions are met, and I will illustrate this better through
the use of an example.

\begin{figure}
\begin{centering}
\includegraphics[scale=0.75]{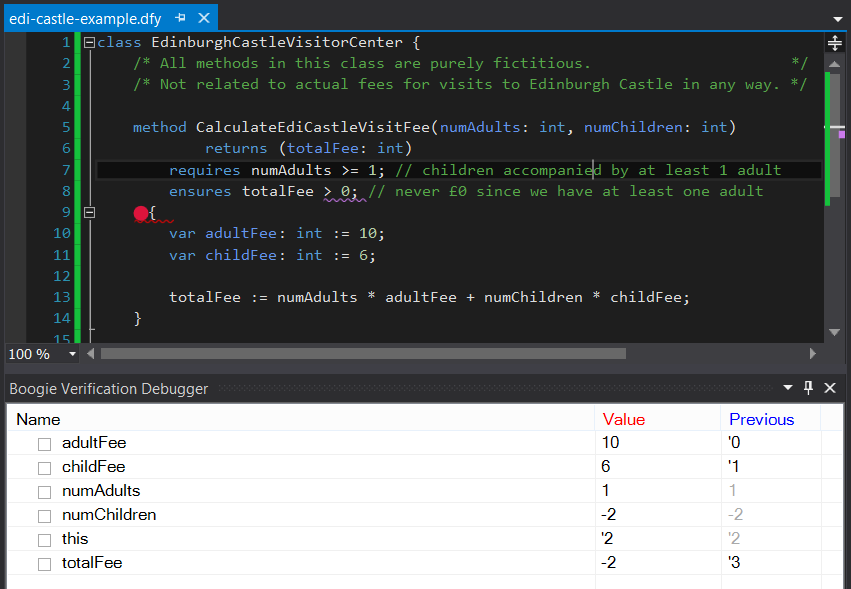}
\par\end{centering}

\protect\caption{Screenshot of Dafny code with the verifier in action, highlighting
a postcondition that does not hold, and showing an example invocation
for which the postcondition fails.\label{fig:Screenshot-of-the}}

\end{figure}

Let's say we would like a method for calculating the cost of a visit
to Edinburgh Castle based on the number of adults and number of children
in a group. As a precondition, we require that children must be accompanied
by at least one adult. Thus, since there will be at least one adult
in the group, we add the postcondition that the total fee will always
be greater than $0$. This fictitious fee calculator can be seen in
Figure \ref{fig:Screenshot-of-the}.%
\footnote{Going forward, the report continues building up on this example by
adding further functionality, and the complete, compilable source
code can be found in the Appendix.%
}

In fact, this Dafny code does not compile, and the debugger provides
us with an example invocation which will cause the postcondition to
fail, with \texttt{numChildren = -2}. This postcondition will in fact
fail whenever the \texttt{numChildren} parameter has a negative value.
There are two ways in which we can fix this: either add another precondition
to ensure the parameter is never less than zero (\texttt{requires
numChildren >= 0}), or change the type of the parameter to \texttt{nat}.

This example, although simplistic, brings out an important distinction
between comments and annotations. Although initially the assumption
of a non-negative fee might have seemed reasonable, it was easy to
miss the case of having a negative number of children, especially
since in real life such a value would not make sense. By putting an
observation about a method in a comment, we have no way of knowing
whether the method actually has this property or not. Not only this,
but when a method is updated we have no guarantee that the corresponding
comments will be updated to match the new functionality. Using annotations,
any code updates causing any postcondition to fail will prevent the
code from compiling until the annotations are updated to reflect the
new code, or, if necessary, the code is fixed to make sure that the
postconditions still hold.

\subsubsection{Assertions\label{sub:Assertions}}

Similar to pre- and postconditions, assertions are also expressions
evaluating to a boolean value. Assertions are however placed in the
middle of a method, and they are used to confirm that ``a particular
expression always holds when control reaches that part of the code''
\cite{koenig2012gettingstarted}.

\begin{lstlisting}[caption={Sample assertions on the method presented in Figure \ref{fig:Screenshot-of-the}.},label={lis:Sample-annotations-on},float,basicstyle={\small\ttfamily},tabsize=4]
var totalFee := CalculateEdiCastleVisitFee(2, 2); // 10 * 2 + 6 * 2 = 32
assert(totalFee > 0); // assertion 1 - verifiable
assert(totalFee == 32); // assertion 2 - non-verifiable
\end{lstlisting}

Let's say I want to create some assertions on the output of the \texttt{CalculateEdiCastleVisitFee()}
method, as seen in Listing \ref{lis:Sample-annotations-on}. As indicated
by the comments, the first assertion can be verified by Dafny, and
this is due to the postcondition which ensures the fee is always greater
than zero. Dafny cannot, however, verify the second assertion, even
though we know this to be true. This is because Dafny's knowledge
of the method is limited to its annotations, as explained in Section
\ref{sub:Methods,-Functions-and}, and it therefore has no way of
knowing that the \texttt{totalFee} value will actually be $32$.

\begin{lstlisting}[caption={A new function method and corresponding, verifiable assertions.},label={lis:fnAssertions},float,basicstyle={\small\ttfamily},tabsize=4]
function method GetDiscountedFamilyTicket(isWeekday: bool) : nat
{
	if (isWeekday) then 22 else 27 	
}

var familyTicketWeekday := GetDiscountedFamilyTicket(true);
assert familyTicketWeekday == 22; // assertion 3 - verifiable
assert GetDiscountedFamilyTicket(false) == 27; // assertion 4 - verifiable
\end{lstlisting}

On the other hand, Dafny can verify both assertions in Listing \ref{lis:fnAssertions},
since it can use the body of the function method for verification.
Note that, as in assertion 4 of Listing\ref{lis:fnAssertions}, functions
can also be used directly in assertions, without having to assign
their return value to to a parameter. This is not possible for methods
since a method can have multiple return values.

\subsubsection{Loop Invariants and Termination Metrics}

Loops pose a problem for Dafny because it is not possible to know
in advance how many times the code inside the loop will be executed,
but the verifier needs to consider all possible paths of code execution.
Loop invariants are another form of annotations available in Dafny,
which enable to the verifier to work with loops. As for the previously
seen annotations, loop invariants are also boolean expressions. Loop
invariants must hold \textit{before} entering the loop, and \textit{for
every execution} of the loop. When trying to prove assertions after
the execution of a loop, Dafny only takes into consideration the loop's
guard (condition), and its invariants. The loop body is not taken
into consideration, similar to how a method's body is ignored when
outside of the method. In other words, when a loop exits, Dafny will
only know that the loop guard failed and that the invariant still
holds.

For example, consider the Dafny method in Listing \ref{lis:audioguides},
which keeps track of the number guides assigned to people in a very
simplistic loop. The method's precondition ensures that there are
enough audio guides for everybody, and thus after loop execution it
should be easy to confirm that the number of assigned audio guides
is equal to the number of people. However, Dafny cannot verify this
assertion because it knows nothing about the body of the loop, or
in other words, Dafny sees the loop body as a black box. After loop
execution, all it knows at the assertion point is that the loop guard
has failed, meaning that \texttt{numAssignedGuides} \texttt{>=} \texttt{numPeople},
but has no way of proving that \texttt{numAssignedGuides} \texttt{==}
\texttt{numPeople}. 

\begin{lstlisting}[caption={A Dafny method which uses a loop to keep track of the number of assigned
audio guides.},label={lis:audioguides},float,basicstyle={\small\ttfamily},tabsize=4]
method AssignAudioGuides(numPeople: nat, numAvailableGuides: nat)
		returns (remainingGuides: nat)
	requires (numAvailableGuides >= numPeople);
{
	var numAssignedGuides := 0;
	while (numAssignedGuides < numPeople)
	{
		numAssignedGuides := numAssignedGuides + 1;
	}

	assert numAssignedGuides == numPeople; // assertion 5 - non-verifiable

	remainingGuides := (numAvailableGuides - numAssignedGuides);
}
\end{lstlisting}

For Dafny to be able to verify the assertion, we add the loop invariant
in Listing \ref{lis:invariant} and this, together with the loop guard,
is enough for Dafny to be able to proof that \texttt{numAssignedGuides}
will contain the correct value after loop execution. Note how the
loop invariant uses \texttt{<=} and not \texttt{<}, like the loop
guard. This is because the loop invariant must hold for \textit{all}
executions of the loop, even for the very last one when the loop condition
fails and the loop exits.

In the Appendix, I also include a \texttt{VerifyAdults()} method which
requires the use of a slightly more complex loop invariant to verify
an assertion. (The assertion makes use an existential quantifier,
which will be introduced in the next section.)

\begin{lstlisting}[caption={The loop invariant required for the verification of assertion 5 in
Listing \ref{lis:audioguides}.},label={lis:invariant},float,basicstyle={\small\ttfamily},tabsize=4]
	while (numAssignedGuides < numPeople)
		invariant numAssignedGuides <= numPeople;
	{
		...
	}
\end{lstlisting}
In the case of loops and recursive functions or methods, Dafny also
proves \textbf{termination}, and this can be done in one of two ways:
either with the use of an explicit \texttt{decreases} annotation,
or by a correct guess from Dafny. For a \texttt{decreases} annotation,
Dafny proves two things: that the expression does indeed get smaller,
and that it is bounded.

Note that the code provided in Listings \ref{lis:audioguides} and
\ref{lis:invariant} does not use a \texttt{decreases} annotation,
but Dafny was still able to verify its termination. This is because
the loop has a common form and thus allows Dafny to guess the correct
termination measure \cite{dafnyTermination}. Alternatively, we could
explicitly add the \texttt{decreases} annotation shown in Listing
\ref{lis:termination}, which is the same as that which will be guessed
by Dafny.

\begin{lstlisting}[caption={The termination measure required for proving that the code in Listings
\ref{lis:audioguides} and \ref{lis:invariant} terminates.},label={lis:termination},float,basicstyle={\small\ttfamily},tabsize=4]
	while (numAssignedGuides < numPeople)
		invariant numAssignedGuides <= numPeople;
		decreases numPeople - numAssignedGuides;
	{
		...
	}
\end{lstlisting}

\subsubsection{Quantifiers, Predicates and Frames}

There are two \textbf{quantifiers} in Dafny: \texttt{forall} and \texttt{exists}.
The \texttt{forall} quantifier corresponds to the universal quantifier
in predicate logic, $\forall$, and can be used to check whether a
property holds for all elements of an array or data structure. The
\texttt{exists} quantifier corresponds to the existential quantifier,
$\exists$, and can be used to check whether a property holds for
at least one element in an array or data structure. A \textbf{predicate}
in Dafny is simply a function that returns a boolean value. Listing
\ref{lis:predicate} shows an example of a predicate whose unary expression
makes use of an existential quantifier. It is important to note that
we would not have been able to create this predicate if Dafny did
not allow for quantification, as we would have had to loop through
the elements of the array one by one, and function bodies can only
contain one expression.

\begin{lstlisting}[caption={A predicate function making use of an existential quantifier.},label={lis:predicate},float,basicstyle={\small\ttfamily},tabsize=4]
	function ChildPresent(visitorsAges: array<int>): bool
	{
		// non-verifiable
		exists i :: 0 <= i < visitorsAges.Length ==> visitorsAges[i] < 18
	}
\end{lstlisting}

\sloppypar{}The code in Listing \ref{lis:predicate}, however, does
not compile. This is due to two reasons. Firstly, \texttt{visitorsAges.Length}
could cause a run-time error if \texttt{visitorsAges} is \texttt{null},
and thus the Dafny verifier is unable to prove the absence of run-time
errors. This can be fixed by adding a function precondition to make
sure that the array is always instantiated.

Secondly, the array access at \texttt{visitorsAges{[}i{]}} causes
the following error: ``insufficient reads clause to read array element''.
This is because we have not included the array in the function's so-called
\textbf{reading frame}. A \texttt{reads} annotation, unlike the other
annotations we have seen so far, is not a boolean expression, but
a set of memory locations that the function has access to. When writing
a function or a predicate, reading frames are important because we
need to know whether a predicate still holds after changes to certain
memory locations are made. So, by adding the \texttt{reads} annotation
for the \texttt{visitorsAges} array, we not only allow the function
to read from the array's memory locations, but also imply that the
predicate might no longer hold if changes to the array are made, which
is what we desire. The compilable version of the \texttt{ChildPresent}
predicate can be seen in Listing \ref{lis:predicateFixed}.

\begin{lstlisting}[caption={An updated, compilable version of the predicate in Listing \ref{lis:predicate}.},label={lis:predicateFixed},float,basicstyle={\small\ttfamily},tabsize=4]
	function ChildPresent(visitorsAges: array<int>): bool
		requires visitorsAges != null; //prevent "target object may be null"
		reads visitorsAges; //prevent "insufficient read clause"
	{
		exists i :: 0 <= i < visitorsAges.Length ==> visitorsAges[i] < 18
	}
\end{lstlisting}

Methods, on the other hand, can read from any memory location they
like, without the need for a \texttt{reads} annotation. However, any
memory locations a method needs to write to must be listed in \texttt{a
modifies} annotation. This annotation, similar to \texttt{reads},
also takes a set of memory locations, and is used to specify a method's
\textbf{writing frame}. The \texttt{read} and \texttt{modifies} clauses
are what enable Dafny's \textit{dynamic frames} \cite{leino2009specandver}.
Through reading and writing frames, arbitrary memory modifications
are limited to something Dafny can reason about, allowing the verifier
to work on one method at a time \cite{koenig2012gettingstarted},
and thus allowing for \textit{modular verification}, meaning that
``separate verification of each part of the program implies the correctness
of the whole program'' \cite{leino2010autoprogver}.

\subsection{From Code to Verification\label{sub:From-Code-to}}

In the previous section we went over some of the main features of
the Dafny language, but we have not yet described how the verifier
goes about generating a mathematical proof of functional correctness.
We will not go over this in detail in this report, but the high-level
process is as follows.

The Dafny verifier actually translates the Dafny code into the intermediate
verification language Boogie 2. The Boogie tool then generates first-order
verification conditions from the intermediate program using the concept
of \textit{weakest preconditions} \cite{leino2009specandver}, and
passes these verification conditions on to the Z3 SMT (satisfiability-modulo-theories)
solver. In this way, correctness of the Boogie program implies correctness
of the original Dafny program \cite{leino2010autoprogver}. Going
back to Figure \ref{fig:Screenshot-of-the}, one can notice that it
is the ``Boogie Verification Debugger'' which presents us with the
counterexample.

SMT-based program verifiers are fully automatic, meaning that they
require no user interaction, and thus this translation approach allows
the functional verification of a Dafny program to occur through the
use of a powerful and state-of-the-art SMT, while still keeping the
interaction required by the programmer limited to the programming
domain \cite{leino2010autoprogver}. This approach, however, still
has its limitation, as the power of the programming language is limited
to the power of the intermediate verification language. For example,
in \cite{leino2010autoprogver} we see that, although Z3 provides
support for algebraic datatypes, Boogie does not, and so there is
no way for Dafny to tap into that support.

\section{Resources and Related Works\label{sec:Resources}}

For anyone wanting a practical introduction to Dafny, I would suggest
having a look at the online tutorial at \url{http://rise4fun.com/Dafny/tutorial/guide}
\cite{getstart_onlne}. The contents are very similar to those of
the paper cited as \cite{koenig2012gettingstarted}, although \cite{koenig2012gettingstarted}
also includes a handy ``Dafny Quick Reference'' appendix. The online
version however has the advantage of having an interactive panel,
allowing the user to try Dafny in the web browser itself, without
the need for any installations. The online tutorial also contains
useful links to other tutorials on language features such as Sets,
Termination \cite{dafnyTermination} and Lemmas.

The lecture notes from the 2008 Marktoberdorf summer school, \cite{leino2009specandver},
are more mathematical-oriented, as they go into the details of Dafny's
logical encoding and the translation from Dafny to Boogie. They are
more focused on explaining how to build a first-order automatic program
verifier, than on how to actually use the Dafny language. Nonetheless,
they give good insight to some of the design decisions which where
taken during the early design stages for Dafny. One must however keep
in mind that a lot of progress has been made on Dafny since 2008,
and some of the statements in \cite{leino2009specandver} no longer
hold. For example, at the time of writing of \cite{leino2009specandver},
Dafny had no support for higher-order functions, but these have now
been included as a new language feature in Dafny 1.9.1 \cite{releasenotes}.
Similarly, Dafny's current type system is very different from that
presented in \cite{leino2009specandver}; it now supports a much larger
set of types, as can be seen from the latest version's type system
documentation at \url{http://research.microsoft.com/en-us/um/people/leino/papers/krml243.html}
\cite{types}.

Another interesting paper is \cite{leino2010autoprogver}. This contains
an overview of some of Dafny's language features, and also includes
a full functional specification of the Schorr-Waite algorithm in Dafny.
Interestingly, in \cite{leino2010autoprogver}, Leino points out that,
at the time of writing, the $120$ lines of Dafny code for the algorithm
were the shortest mechanical-verifier input to date, with the closest,
shortest implementation being 400 lines of Isabelle proof scripts.
It was also the first Schorr-Waite proof to be carried out solely
by an SMT-solver, and verfication took only 5 seconds. However, Leino
himself, who is the principal researcher behind Dafny, admits in \cite{leino2010autoprogver}
that coming up with the proof was not an easy task, especially considering
the ``mouthful'' of requried invariants, and concludes that the
``task is not yet for non-experts'' \cite{leino2010autoprogver}.
This, in my opinion is one of Dafny's limitations, as it indicates
that it is not always the case that writing bug-free annotations is
easier than writing bug-free code.

In \cite{leino2010autoprogver}, Leino also gives a good reference
to a number of related works, namely specification languages, and
how they compare to Dafny. These include the Java Modeling Language
(JML), Spec\# and VeriCool 1. JML and Spec\# both make use of \textit{pure
methods} instead of mathematical functions, and Leino indicates that
these are much harder to get right, going as far as to include the
slogan ``pure methods are hard, functions are easy'' \cite{leino2010autoprogver}.
VeriCool 1 inspired the use of dynamic frames in Dafny, and the languages
share various similarities, but differ in the way ghost variables
are handled, where Dafny's specification avoids problems with recursive
functions \cite{leino2010autoprogver}.

\section{Conclusion\label{sec:Conclusion}}

Dafny originally started out as an exercise in encoding dynamic frames,
and has now developed into a general-purpose programming language
and a static verifier for functional correctness \cite{leino2010autoprogver}.

Personally, I think one of the main advantages of the Dafny verifier
is that the programmer can interact with it in much the same way as
with the compiler. The Dafny language syntax itself is not difficult
to get used to, as it is quite similar to other languages, such as
Java and C\#. Also, through ghosting, one can include verification
code without affecting the performance of the executable program itself.
Another advantage is that Dafny is both concise and fast, as shown
for example through the implementation of the Schorr-Waite algorithm
in \cite{leino2010autoprogver}.

On the other hand, some limitations of Dafny include the fact that
its support is limited to what can be represented in the intermediate
verification language Boogie 2. Also, while coming up with the required
annotations for toy examples is easy, this process gets more and more
difficult as the code gets more complex, and this was also demonstrated
in \cite{leino2010autoprogver}. Also, the fact that, for example,
support for ``string'' and ``char'' types has just been added
in the latest version, indicates that there is still a long way to
go if Dafny is to be used for the implementation of large-scale systems.

However, a significant number of features have already been added
to the original Dafny version presented in the original lecture notes
\cite{leino2009specandver}. Development of the language and verifier
is still active and ongoing, as shown by the fact that the latest
version was released just over two weeks ago in October 2014; and,
all in all, I believe that Dafny has proved itself to be a very promising
tool for the automatic, statical verification of full functional correctness
of programming code. 

\bibliographystyle{plain}
\bibliography{references}

\begin{thebibliography}{1}

\bibitem{leino2010autoprogver}
K~Rustan~M Leino.
\newblock {Dafny: An Automatic Program Verifier for Functional Correctness}.
\newblock In {\em Logic for Programming, Artificial Intelligence, and
  Reasoning}, pages 348--370. Springer, 2010.

\bibitem{koenig2012gettingstarted}
Jason Koenig and K~Rustan~M Leino.
\newblock {Getting Started with Dafny: A guide}.
\newblock {\em Software Safety and Security: Tools for Analysis and
  Verification}, 33:152--181, 2012.

\bibitem{codeplex}
K~Rustan~M Leino.
\newblock {CodePlex - Dafny: An Automatic Program Verifier for Functional
  Correctness}.
\newblock \url{http://dafny.codeplex.com/}, 2014.
\newblock [Online; accessed 02-Nov-2014].

\bibitem{softFailures}
Gang Tan.
\newblock {A Collection of Well-Known Software Failures}.
\newblock \url{http://www.cse.lehigh.edu/~gtan/bug/softwarebug.html}, 2009.
\newblock [Online; accessed 04-Nov-2014].

\bibitem{releasenotes}
K~Rustan~M Leino.
\newblock {CodePlex - Dafny 1.9.1 Release Notes}.
\newblock \url{http://dafny.codeplex.com/releases/view/135602}, 2014.
\newblock [Online; accessed 04-Nov-2014].

\bibitem{dafnyTermination}
Microsoft Research.
\newblock {rise4fun - Dafny - Termination}.
\newblock \url{http://rise4fun.com/Dafny/tutorial/Termination}, 2014.
\newblock [Online; accessed 05-Nov-2014].

\bibitem{leino2009specandver}
K~Rustan~M Leino.
\newblock {Specification and Verification of Object-Oriented Software}.
\newblock {\em Engineering Methods and Tools for Software Safety and Security},
  22:231--266, 2009.

\bibitem{getstart_onlne}
Jason Koenig and K~Rustan~M Leino.
\newblock {rise4fun - Getting started with Dafny: A guide}.
\newblock \url{http://rise4fun.com/Dafny/tutorial/guide}, 2014.
\newblock [Online; accessed 06-Nov-2014].

\bibitem{types}
K~Rustan~M Leino.
\newblock {Microsoft Research - Types in Dafny}.
\newblock
  \url{http://research.microsoft.com/en-us/um/people/leino/papers/krml243.html},
  2014.
\newblock [Online; accessed 06-Nov-2014].

\end{thebibliography}
\pagebreak{}

\appendix

\section*{Appendix\label{sec:Appendix-A}}

\begin{lstlisting}[basicstyle={\small\ttfamily},breaklines=true,tabsize=4,frame=none]
class EdinburghCastleVisitorCenter {
	/* All methods in this class are purely fictitious.						 */
	/* Not related to actual fees for visits to Edinburgh Castle in any way. */

	method CalculateEdiCastleVisitFee(numAdults: nat, numChildren: nat)
			returns (totalFee: int)
		requires numAdults >= 1; // children accompanied by at least 1 adult
		ensures totalFee > 0; // never £0 since we have at least one adult
	{
		var adultFee: int := 10;
		var childFee: int := 6;

		totalFee := numAdults * adultFee + numChildren * childFee;
	}

	function method GetDiscountedFamilyTicket(isWeekday: bool) : nat
	{
		if isWeekday then 22 else 27
	}

	method FamilyTicketVerification()
	{
		var numAdults := 2; // type inference
		var numChildren := 2; 

		var totalFee := CalculateEdiCastleVisitFee(numAdults, numChildren);
		// totalFee = 10 * 2 + 6 * 2 = 32

		assert(totalFee > 0); // possible because of postcondition
		//assert(totalFee == 32); // cannot verify, no related postcondition

		var familyTicketWeekday := GetDiscountedFamilyTicket(true);
		assert familyTicketWeekday == 22;

		// function can also use directly in annotations
		assert GetDiscountedFamilyTicket(false) == 27; 
	}

	method AssignAudioGuides(numPeople: nat, numAvailableGuides: nat)
			returns (remainingGuides: nat)
		requires (numAvailableGuides >= numPeople);
	{
		var numAssignedGuides := 0;
		while (numAssignedGuides < numPeople)
			invariant numAssignedGuides <= numPeople;
			decreases numPeople - numAssignedGuides; // not necessary (Dafny can guess)
		{
			numAssignedGuides := numAssignedGuides + 1;
		}

		assert numAssignedGuides == numPeople; // fails without invariant

		remainingGuides := (numAvailableGuides - numAssignedGuides);
	}

	method VerifyAdults(adultAges: array<int>) returns (allAdults: bool)
		requires (adultAges != null); // cannot access "Length" without this
	{
		var index := 0;

		while (index < adultAges.Length)
			decreases (adultAges.Length - index);
			invariant index <= adultAges.Length;
			invariant forall i :: 0 <= i < index ==> adultAges[i] >= 18;
			// without the last invariant,
			// Dafny has no way of knowing we checked all "previous" values
		{
			if (adultAges[index] < 18)
			{
				allAdults := false;
				break;
			}

			index := index + 1;
		}

		if (allAdults)
		{
			assert forall i :: 0 <= i < adultAges.Length ==> adultAges[i] >= 18; // fails without "forall" invariant
		}
		
	}

	// predicate
	function ChildPresent(visitorsAges: array<int>): bool
		requires visitorsAges != null; // prevent "target object may be null"
		reads visitorsAges; // prevent "insufficient read clause"
	{
		exists i :: 0 <= i < visitorsAges.Length ==> visitorsAges[i] < 18
	}

}
\end{lstlisting}

\end{document}